\documentstyle[12pt]{article}

\def \be {\begin{equation}}
\def \ee {\end{equation}}
\def \bea {\begin{eqnarray}}
\def \eea {\end{eqnarray}}

\begin{document}

\begin{flushright}
ITP-SB-97-41\\
LPTHE-ORSAY 97/40
\end{flushright}

\vspace{3mm}
\begin{center}
{\Large \bf Power Corrections\\ and Nonlocal Operators\footnote{Presented
at the Fifth International Workshop on Deep Inelastic
Scattering and QCD, Chicago, IL, April 14-18, 1997.}} 
\end{center}
\vspace{2mm}
\begin{center}
Gregory P.\ Korchemsky\\
\vspace{2mm}
{\it Laboratoire de Physique Th\'eorique et Hautes Energies\\
Universit\'e de Paris XI, Centre d'Orsay\\
91405 Orsay C\'edex, France} \\
\vspace{4mm}
Gianluca Oderda and George Sterman\\
\vspace{2mm}
{\it Institute for Theoretical Physics\\
State University of New York at Stony Brook\\
Stony Brook, NY 11794-3840, USA} \\  
\vspace{2mm}
April 1997
\end{center}

\begin{abstract}
We discuss power corrections to infrared safe cross sections
and event shapes, and identify
a nonperturbative function that governs $1/Q$ corrections
to these quantities.
\end{abstract}

\section{Introduction}

Nonperturbative corrections to infrared safe
 jet cross sections and event shapes are
an important issue in the study of QCD.
They are the natural starting point for
a unified perturbative-nonperturbative treatment of 
high energy cross sections, because
they enter at the level of nonleading powers
of the hard momentum scale, $Q$.

For example, a classic analysis  \cite{Mueller}
of the total cross section for ${\rm e}^+{\rm e}^-$ annihilation
to hadrons identifies terms
$\alpha_s^n(Q^2)b_2^n\; n!$ at $n$th order, which may be
attributed to the infrared (IR) behavior of the QCD running coupling,
$\alpha_s(k^2)=4\pi/[b_2\ln(k^2/\Lambda^2)]$.  
Borel analysis shows that this nonconvergence is 
due to an ambiguity in the cross section at
$Q^{-4}$ relative to leading behavior.
This ambiguity is in direct correspondence
with the contribution of the gluon condensate to the operator
product expansion (OPE).  The OPE, however, is
available in only a few cases.
Nevertheless, we would like to abstract from these considerations  a
method of ``substitution".  That is, we shall assume that the contribution of 
{\em any} region of momentum
space where the contour integrals of perturbation theory (PT) are trapped by mass-shell
and IR
singularities is a source of nonperturbative corrections,
whose power suppression may be estimated from PT itself.
We outline a general procedure, which we
shall briefly illustrate below.

Let $\sigma$ an be an IR safe cross section at large
scale $Q$: 
{(i)} Identify regions $R$ in momentum space where lines are
pinched on-shell in $\sigma$, by use of Landau equations or
an equivalent method.
{(ii)} Organize logarithms of momenta $k$ that occur
in $R$ into: (a) $\alpha_s\left(f(k)\right)$,
with $f(k)$ a characteristic momentum scale, and/or 
(b) explicit kinematic
integrals. 
{(iii)} Introduce a cutoff $\kappa$ on 
(some) components: $k^\nu<\kappa$, to
define the contribution $\sigma_R$ from region
$R$, such that $\alpha_s\left(f(k)\right)>
\alpha_0$, $\alpha_0$ fixed.
{(iv)} With the coupling fixed, evaluate the power behavior
$\sigma_R \sim Q^{-2-m}\kappa^m$. 
{(v)} Find a ``universal" matrix element 
$\langle {\cal O} \rangle$,  of
dimension $m$, whose perturbative expansion is identical to that for $R$.
{(vi)} Remove $\sigma_R$ from $\sigma$, replacing it
with $\langle {\cal O} \rangle$,
$\sigma= \sigma^{({\rm reg})}(\kappa)
+{\langle {\cal O} \rangle(\kappa)/ Q^{-2-m}}\, .
$
Of these steps, items (ii) and (v) require spectial treatment
on a case-by-case basis.  Nevertheless, this
approach includes the analysis of
infrared renormalons \cite{universal,KoSt}, and represents, we believe, a
somewhat more general viewpoint.
\bigskip

\section{Power Corrections in Event Shapes}

As an application, we consider the perturbative expansion
for infrared safe event shapes, such as the thrust $T$, close to $T=1$,
the limit of two perfectly narrow jets.  The leading behavior for a large class of
such event shapes $w$ is $1/w$ (times logs of $w$) in this limit.  Keeping only the
$1/w$ terms, the program outlined above may be carried out explicitly.
To be specific, we shall assume that to the power $1/w$,
the weight factorizes into contributions from individual
particles,
\be
w(k) = \sum_{\rm particles\ {\it i}} w(k_i/Q)=\sum_{\rm particles\ {\it i}} {k_i^0\over Q}f_w(cos\; \theta_i)\, ,
\label{wsum}
\ee
for some function $f_w$ of the cosine of the angle of particle $i$ to 
the two-jet axis.  In the case of thrust, $f_{1-T}=(1-|cos\; \theta_i|)$.

In the two-jet limit, the differential cross section $d\sigma/dw$
factorizes into functions describing the internal evolutions of
the jets convoluted in $w$ with a 
``two-jet" soft-gluon function $\sigma_2(Q,w)$.
$\sigma_2$ is
the eikonal approximation to
the cross section at fixed weight $w$, in which the jets
are replaced by oppositely-moving, lightlike Wilson lines \cite{KoSt}.
For the $1/w$ contributions, the directions of the Wilson
lines may be considered as fixed.  We now
apply our reasoning to $\sigma_2$.  Suitably defined, the jet
functions give nonleading power corrections.

Our first observation is that $\sigma_2(Q,w)$ is a convolution, 
\be
\sigma_{2}(Q,w) = \sum_{n=0}^\infty {1\over n!}\;
\int dw_1\dots dw_n\; \delta\left(w-\sum_{i=1}^n w_i\right)
\prod_{i=1}^n\; S(Q,w_i)\, ,
\label{sig2Jconv}
\ee
in terms of
a kernel, $S(Q,w)$.  The convolution fixes the weights
of final states that contribute to $\sigma_2(Q,w)$.  Then, if the weight function satisfies
Eq.\ (\ref{wsum}),
the Laplace transform of $\sigma_2(Q,w)$ exponentiates (up to small corrections, which
we suppress),
\be
{\tilde \sigma}_{2}(Q,N)
=
\int_0^{w_{\rm max}} dw\; e^{-Nw}{\sigma_{2}(Q,w)}=\exp[{\tilde S}(Q,N)]\, .
\label{sigexp}
\ee
By construction of $w$, this transform 
is infrared finite.
The perturbative expansion of the kernel $S$ in Eq.\ (\ref{sig2Jconv})
is identical to a sum of two-particle
irreducible diagrams called  ``webs" long ago \cite{GFT}.
Webs have a number of important properties.  First, they
require only a single, additive UV renormalization, corresponding to
multiplicative renormalization for $\sigma_{2}$.  Second, they give rise
to only a single overall collinear and infrared logarithm each, aside
from logarithms which may be organized into the running of the coupling.
These conditions are summarized in the integral representation,
\begin{eqnarray}
{\tilde S}(Q,N) 
&=&
 \int_0^{Q^2}{dk^2\over k^2}
\int_0^{Q^2-k^2} {dk_T^2\over k^2+k_T^2}\; 
\nonumber \\
&\ & \quad \times
\int_{\sqrt{k^2+k_T^2}}^Q{dk_0\over \sqrt{k_0^2-k^2-k_T^2}}\;
\gamma_w \left ( {k\over \mu},{\mu\over Q},\alpha_s(\mu),N\right )\, ,
\label{stilde}
\end{eqnarray}
where $k$ represents $k_0$, $k_T$ and $\sqrt{k^2}$.
The combination ${\gamma_w(k,N)/[k^2(k^2+k_T^2)]}$ is an
integrable distribution for $k^2,k_T^2\rightarrow 0$.
The two overall logarithms of $N$ are generated from the
$k$ integrals.
In Eq.\ (\ref{stilde}), we have implemented items (i) and (ii) in the 
method of substitution above.  To identify the explicit forms of power corrections in $Q$,
we expand $\gamma_w$ in $Q$ at fixed values of $N$.
The precise form of dependence on soft regions in (\ref{stilde}), and
the corresponding substitutions (see above), depend on the 
weight, but are simplified by the web 
structure, leading to clear sources of power corrections.

For $w=1-T$, the expansion of $\gamma_{1-T}$ gives
\be
{\gamma_{1-T}^{(1)}\over k^2}= {N\over Q}\; \delta(k^2)\; 2C_F
{\alpha_s(k_T)\over \pi}\; \left(k_0-\sqrt{k_0^2-k_T^2}\right)
+{\cal O}(N^2/Q^2)\, ,
\label{gammaexpand}
\ee
where we have used the renormalization-group invariance of $\gamma_{1-T}$ to
set $\mu=k_T$.
After the $k_0$ integral in (\ref{stilde}), we find an exponentiating $1/Q$
correction, which multiplies the (infrared regulated) perturbative result,
\be
{\tilde \sigma}_{2}^{(1-T,{\rm corr})}(Q,N) \sim
\exp \left[ {N\over Q}\; {2C_F\over \pi}\;
\int_0^{\kappa} {dk_T}\; \alpha_s(k_T) \right]\, ,
\label{exp1overQ}
\ee
where we have introduced, as above, a new scale $\kappa$ to
isolate the infrared-sensitive region around $k_T=0$.
This, of course, is a typical infrared renormalon, now in the exponent \cite{universal,KoSt}.

We now ask if it is possible to give an operator interpretation
to this result, and thus to ``substitute" 
a nonperturbative matrix element for the IR region
of PT, leading to a universal quantity that controls $1/Q$ behavior  \cite{universal,KoSt}.
To construct this matrix element, we define
operators that measure the energy
that arrives over time at a sphere ``at infinity", 
\be
\Theta(\hat{y})
=
\lim_{|{\vec y}|\rightarrow\infty}\int_0^{\infty}
{dy_0\over (2\pi)^2}\; |\vec{y}|^2\; 
\hat{y}_i\; \theta_{0i}(y^\mu)\, ,
\label{thetadef}
\ee
with $\theta_{\mu\nu}$ the energy-momentum tensor,
and $\hat{y}$ a unit vector.
The energy density that flows in direction $\hat{y}$
for $\sigma_{2}(Q,w)$ is 
\be
{\cal E}(\hat{y})
=
\langle0\mid W^\dagger_{v_1v_2}(0)\; \Theta(\hat{y})\; W_{v_1v_2}(0)\mid 0\rangle\, ,
\label{calEdef}
\ee
where $W_{v_1v_2}(0)$ is the product of outgoing Wilson  lines in the $v_i$
directions, joined by a color singlet vertex at the origin.
In these terms, the leading power correction, Eq.\ (\ref{exp1overQ}), is of the general form
\be
\ln{\tilde \sigma}_{2}^{(w,{\rm corr})}(Q,N) =
{N\over Q}\int {d\Omega_y\over 2\pi}\; 
f_w(\cos\theta)\; {\cal E}(\hat{y})\, ,
\ee
with $f_w$ the function in Eq. (\ref{wsum}).
$1/Q$ corrections for a wide class of event shapes are thus
generated from the nonperturbative function  ${\cal E}(\hat{y})$,
the matrix element of a nonlocal operator,
which describes the nonperturbative component of energy flow, associated with two-jet
color flow.  Generalizations to multijet cross sections
are possible.
We anticipate that this approach will help to unify the treatments
of power corrections for a variety of infrared-safe quantities.

{\em Acknowledgement}.
This work was supported in part by the National Science Foundation,
under grant PHY9309888.  We would like to thank R.\ Akhoury,
Eric Laenen and Nikolaos Kidonakis for helpful conversations.

\end{document}